# Deformation mechanisms of Inconel-718 at nanoscale by molecular dynamics

Abrar Faiyad[1], Md. Adnan Mahathir Munshi [1], Sourav Saha[1,2,*]

[1]Department of Mechanical Engineering, Bangladesh University of Engineering and Technology, Dhaka-1000, Bangladesh.

[2]Theoretical and Applied Mechanics, Northwestern University, Evanston, Illinois, USA.

**Abstract**

Ni-based super alloy Inconel-718 is ubiquitous in metal 3D printing where high cooling rate and thermal gradient are present. These manufacturing conditions are conducive to high initial dislocation density and porosity or void in the material. This work proposes a molecular dynamics (MD) analysis method that can examine the role of dislocations, cooling rates, void, and their interactions governing the material properties and failure mechanism in Incone-718 using Embedded Atom Method (EAM) potential. Three different structures: nanowire (NW), nanopillar, and nanoplate are used throughout this work. Initially, strain rates are varied from $10^8$ s$^{-1}$ to $10^{10}$ s$^{-1}$ keeping the NW diameter and temperature constant at 3.17 nm and 300K respectively. Compressive loading is applied to a 7.04 nm nanopillar by applying a constant strain rate of $10^9$ s$^{-1}$. While, temperature is varied from 100K to 700K with increment of 200K. Finally, different cooling rates ranging from $0.5 \times 10^{10}$ K/s to $1 \times 10^{14}$ K/s are applied to nanoplates (with and without a central void). The size of the central void is kept fixed at 2.12 nm. Our simulations results suggest that Shockley dislocations play key role in plastic deformation under tensile loading. Increasing strain rates in tension not only results in strain hardening but also increase in dislocation density. Our computational method is successful to capture extensive sliding on {111} shear plane, which leads to significant necking of the alloy before fracture due to dislocation. The high cooling rates creating non-equilibrium structure leads to a high strength and ductile behavior. On the other hand, the low cooling rate forming well defined crystalline structure causes low strength and brittle behavior. This brittle to ductile transition is observed solely due to cooling rate. Moreover, cooling rate may diminish the void by healing the structure during solidifications process. Subsequent mechanical properties by varying temperature and size are also presented in detail. This result details a pathway to design parts with Inconel-718 alloy efficiently.

## 1. Introduction

Nickel based super alloys have been the center of extensive research and development and established its place in various crucial applications in the last four decades [1–3]. Inconel-718 is a Ni-Cr-Fe-based austenitic superalloy widely used as a core element for elevated temperature applications such as gas turbines and aerospace industries due to its high-temperature strength (results of solid-solution strengthening and precipitation strengthening), low thermal conductivity, high corrosion resistance, great fatigue resistance and creep resistance [3–7]. However, its high hardness and low thermal conductivity make it unsuitable for most conventional machining processes due to tool-wear and poor workpiece integrity [8–12] which limits the applicability. With the rise of different additive manufacturing methods (AM) like selective laser melting (SLM), structural parts with complex shape can now be made. AM method helps to eliminate the problems encountered during the machining processes for Inconel-718 [1,5,13,14]. As a result, the demand of a material with elevated mechanical properties like Inconel-718 has risen manifolds [6]. Unfortunately, AM comes with its own set of limitations. The reliability and manufacturability of structural parts made by AM depend on the structural properties at the nano- and micro-scale, which, due to small sizes, are expected to be different from bulk counterpart. The mechanical performance depends on the nanoscale micro structural constituents such as dislocations and process induced voids [15–17]. Nanostructures like nanowire, nanopillar, and nanoplate can provide a useful test-bed to analyze the deformation behavior of materials at the nano-scale [18]. Such analyses are important to deepen our understanding on the dynamics of crystal defects such as dislocations and void. Molecular dynamics can be a useful tool to observe the physics of deformation mechanism at this scale.

For Inconel-718, the machinability properties of different phases prevailing at different temperatures have been researched extensively [8,9,19–21]. However, SLM holds the key to efficient and intricate designing and manufacturing of Inconel-718 parts. So, thermal and mechanical properties of SLM manufactured Inconel-718 parts have also been explored previously in great details [1,4–7,14,22–24]. Besides, there is a number of computational studies including effects of oil mist spraying on the Inconel-718 parts, ploughing effects of cutting-edge scale, growth of creep cavities, crystal plasticity based computations on modeled and AM Inconel-718, numerical determination of heat effected areas, and determination of density of normal and metastable Inconel-718 [12,24–35]. Despite the wealth of studies on different behavior and properties of Inconel-718 using experiments and in multiscale modeling, evidently no study has been performed to provide an insight into the physics of deformation mechanism of the material at the nanoscale. Moreover, the effect as-built dislocation density and dynamics in Inconel-718 are yet to be revealed in the context of additive manufacturing. The cooling rate of the printed parts are of unparallel importance in controlling the porosity and defects in the finished parts produced by different additive manufacturing techniques [36–38]. Hence it is also necessary to understand the effects of cooling rate and void on material properties and deformation mechanism at nanoscale clearly.

Keeping this scope in mind, this paper presents atomistic simulation results of single-crystalline (Sub-grain) Inconel-718 nanostructures under uniaxial tensile and compressive load. Nanostructures such as nanowires, nanopillar, and nanoplate are used to gain comprehensive insights throughout this study. Even with the advancements of additive manufacturing, the manufacturing of extremely small structures: nanowires (NW), nanopillars (NP), and nanoplate are not far out of reach.[39,40] . The effects of strain rate on NW and nanopillar is investigated to give proper insights about dislocations. The impact of different cooling rates on nanoplates with and without void is also observed. Moreover, effects of temperature and size on the mechanical properties of Inconel-718 NWs are investigated. Failure mechanism is elucidated to explain the failure behavior of the Inconel-718 NWs. This study will help not only to obtain a better understanding of this material behavior in nanoscale but also help better design of AM materials.

## 2. Methodology

*2.1 Atomic Structure*

Inconel-718 is mainly composed of 3 major elements Ni, Cr and Fe which accounts almost 91% of the alloy. The remaining trace elements are Cu, Mo, Nb, C, Mn, P, S, Si, Ti, Al and Co forms not more than 9% of the composition [41]. For comprehensive understanding of its actual chemical composition, proportions of different materials are given in Table 1.

Table 1: Chemical composition (%) of Inconel – 718[41]

| Ni | Fe | Cr | Cu | Mo | Nb | C | Mn | P | S | Si | Ti | Al | Co |
|---|---|---|---|---|---|---|---|---|---|---|---|---|---|
| 50-55 | Remainder | 17-21 | .30 max | 2.8-3.3 | 4.75-5.50 | .08 max | .35 max | .015 max | .015 max | .35 max | .65-1.15 | .2-.8 | 1.0 max |

Modeling nano-structure of Inconel-718 using all the elements is onerous. The problem is the redistribution of the other elements and modeling their interatomic interactions would be difficult. However, study performed by Wang *et al*[28] showed that the density function of simplified FCC structure of $Ni_{60}Cr_{21}Fe_{19}$ matched well to the bulk Inconel-718. While preparing the model, Wang *et al* replaced the trace elements with different percentages of the major elements (Ni, Cr, Fe). In this study, the same strategy is used to generate the geometry.

In all the simulations performed, the lattice is first made by creating an FCC structure made of Ni atoms with [0 0 1] crystal orientation. After that 21% and 19% of all the Ni atoms are replaced randomly by Cr and Fe atoms respectively. To have complete insights into the mechanical properties, three different nanostructures (nanowire, nanopillar, and nanoplate) are used throughout this study. The aspect ratio of height to width is kept constant as 10:1 for nanowire, 2:1 for nanopillar. But, for the square nanoplate, length to thickness ratio is 10:1. As a representation of the Inconel-718, prepared nanowire and nanoplate (containing a central void) models are shown in figure 1. Here, L, W, and H stand for the Length, Width, and Height respectively**.** Moreover, Figure 1(c) represents lattice structure for single crystal of Inconel-718 where *Cr* and *Fe* seen to be distributed randomly.

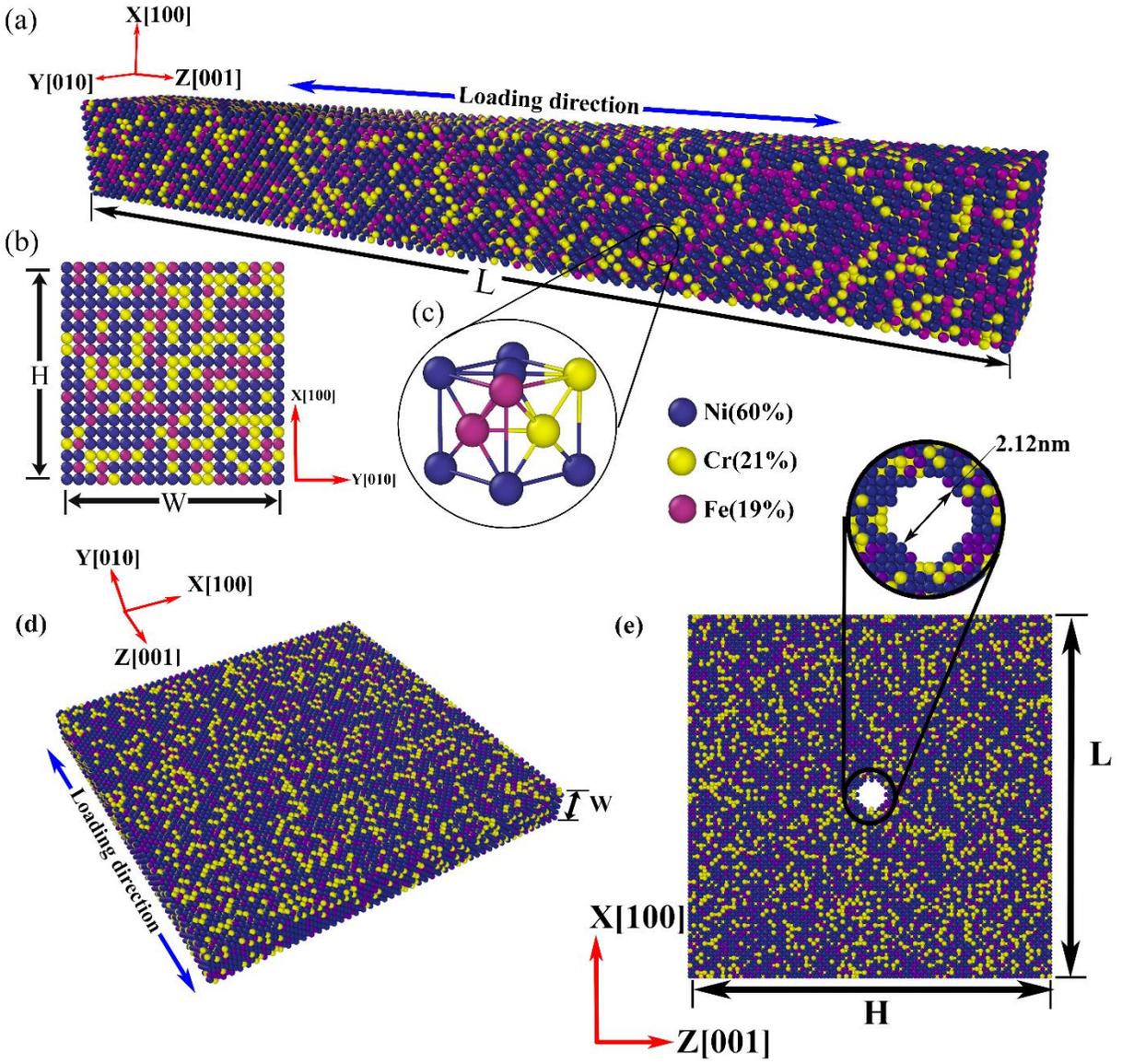

Figure 1: (a) The initial co-ordinates of Inconel-718 nanowire where (b) cross section perpendicular to the Z direction is shown. (c) Unit cell showing the random distribution of Cr and Fe atoms in the Ni lattice (d) The initial co-ordinates of the nanoplate with (e) cross section of the nanoplate perpendicular to the Y direction with a better view of the central void.

For performing temperature, size, and strain rate dependency simulations, nanowire structure is used. In all cases, the periodic boundary condition is applied along the loading axis of nanowire and the remaining surfaces are free surface. Additionally, to understand dislocation dependent behavior clearly, nanopillar is used. Finally, a square nanoplate with periodic boundary in all three axes, is used to determine the effects of cooling on the mechanical properties of Inconel-718. For a better understanding of alloy and void interaction at nanoscale, a central void is introduced to the plate and then effects of annealing on the void is observed. The geometric attributes of Inconel-718 nanostructures are presented in Table 2.

Table 2: Geometric attributes of Inconel-718 nanostructures

|  | Width (nm) | Length (nm) | Surface area in direction of strain (nm$^2$) | Number of atoms |
|---|---|---|---|---|
| Nanowire | 2.47 | 24.65 | 6.08 | 15750 |
|  | 2.81 | 28.13 | 7.91 | 23120 |
|  | 3.17 | 31.7 | 10.05 | 32490 |
|  | 5.28 | 52.83 | 27.91 | 144150 |
|  | 5.99 | 59.87 | 35.85 | 208250 |
|  | 6.69 | 66.92 | 44.78 | 288990 |
| Nanopillar | 7.04 | 14.09 | 49.62 | 131351 |
| Nanoplate | 1.76 | 17.61 | 31.01 | 50000 |

*2.2 Computational Details*

All the simulations are carried out using the Large-scale Atomic/Molecular Massively Parallel Simulator (LAMMPS) software package[42], and OVITO[43] is used for visualization of atomistic deformation processes. The choice of interatomic potential has a significant impact on molecular simulations. In this work, we use the embedded atomic model (EAM) potential and parameters [44,45] which was firstly introduced by Daw and Baskes and employed successfully to create the MD simulation model of Ni-Cr-Fe tertiary alloy [46,47]. The time step for all simulations are taken as 1fs. Before applying tensile and compressive load to NWs and nanopillar, NVE is performed first for 20 ns. Then isothermal-isobaric (NPT) ensemble is applied for pressure equilibration at atmospheric pressure and prescribed temperature for 20 ns. Finally, the system is thermally equilibrated by canonical (NVT) ensemble for 20 ns. In order to control the temperature, a Nose-Hoover thermostat is employed in these steps. Moreover, to equilibrate various state variables, the timesteps mentioned are chosen by trial and error for NVE, NPT, and NVT simulations. Temperature is varied from 100K to 800K by applying a constant strain rate of $10^9$ s$^{-1}$ to a 3.17 nm nanowire. Then, the effects of different strain rates are observed by performing the simulation at $10^8$ s$^{-1}$ to $10^{10}$ s$^{-1}$ strain rates on the same nanowire while maintaining 300K constant temperature. Such a high strain rates allow to perform simulations with reasonable computational resource, and have been commonly applied in atomistic scale simulations to investigate the material failure phenomena [48–50]. For comprehensive study of the dislocation, a nanopillar is used. Finally, a nanoplate is used to determine the effects of cooling rates. The nanoplate is at first heated to 2000K temperature at a rate of $10^{12}$ K/s. After that the plate is equilibrated for 0.1 ns at 2000K temperature to ensure even melting of the structure. Following this the plate is cooled to 300K temperature at 0.5x10$^{10}$ K/s, 1x10$^{10}$ K/s, 2x10$^{10}$ K/s, 1x10$^{11}$ K/s, 1x10$^{12}$ K/s, 1x10$^{13}$ K/s cooling rates. Lastly, to observe the effects of void on the fracture dynamics of the plate, a central void of 2.12nm is introduced to the plate and a comparative study of the nanoplate with and without annealing is performed.

The stress-strain relation of all the structures are studied by applying a strain at a rate of $10^9$ s$^{-1}$ on all the plates along their length. In all cases, the viral stresses are calculated using the formula.

$$\sigma_{virial}(r) = \frac{1}{\Omega}\sum_i [(-m_i \dot{u}_i \otimes \dot{u}_i + \frac{1}{2}\sum_{j \neq i} r_{ij} \otimes f_{ij})] \tag{1}$$

where the sum is taken for all the atoms in the volume, $m_i$ denotes the mass of atom $i$, $\dot{u}_i$ is the time derivative of the displacement, $r_{ij}$ denotes the position vector and $f_{ij}$ denotes the interatomic force applied on atom $i$ by atom $j$.

## 3. Validation

The density function of the modeled lattice of $Ni_{60}Cr_{21}Fe_{19}$ was reported to have a close proximation to the actual density of Inconel-718 at different temperatures in previous works.[28] Further tests are performed to validated both the structure and interatomic potential . For this purpose, two cubes are prepared (as presented in table 3) and validated for the Young's modulus (E) and melting point of the cubes with respect to the experimentally available data.

Table 3: Geometric attributes of cubes for method validation

|   | No. of atoms | Boundary condition |
|---|---|---|
| 1 | 13720 | Periodic in all directions |
| 2 | 34461 | Free surface in all direction |

We apply a strain rate of $10^9$ s$^{-1}$ in the first cube at temperatures ranging from 100K to 800K and determine the Young's modulus (YM) of the material throughout the range and compare it with the experimentally available data from the Special Metal Corporation®[51]. It is observed from figure 2 that the experimental YM [52] closely resembles the Young's modulus of the present study.

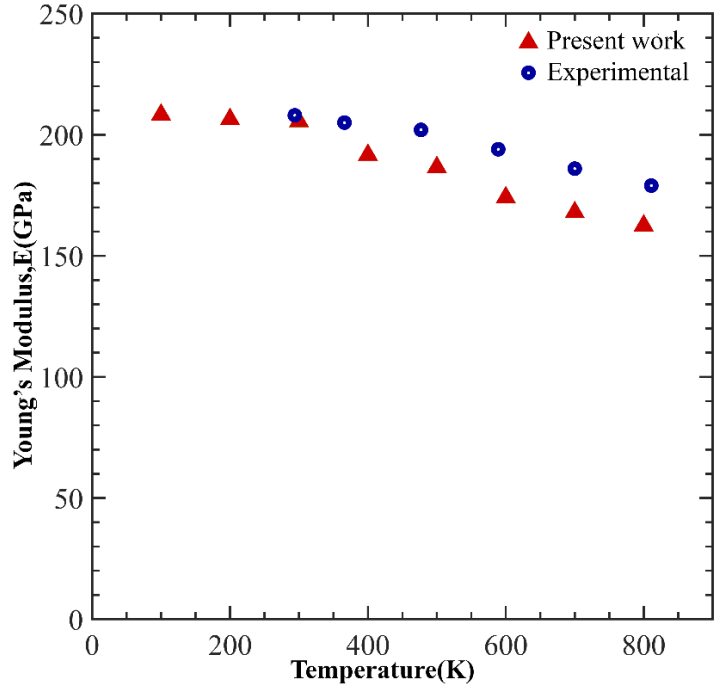

Figure 2: Comparison of Young's modulus obtained in the current study with available experimental values [52].

For the second cube, the temperature is increased from 300K to 3000K and resulting change in the total energy of the cube with change of temperature is plotted. It is observed that there is a steep change in the slope of the curve indicating the initiation of bond breaking due to phase change of the structure and again a second change of slope which signifies that the cube has completely melted (figure 3). It is observed that the material is completely melted within a range of 1621K-1705K which is in good agreement with the experimentally available melting range of 1643K-1703K. The proximity of the values obtained by both the method verifies the present method and adopted potential to describe the interactions in Inconel-718.

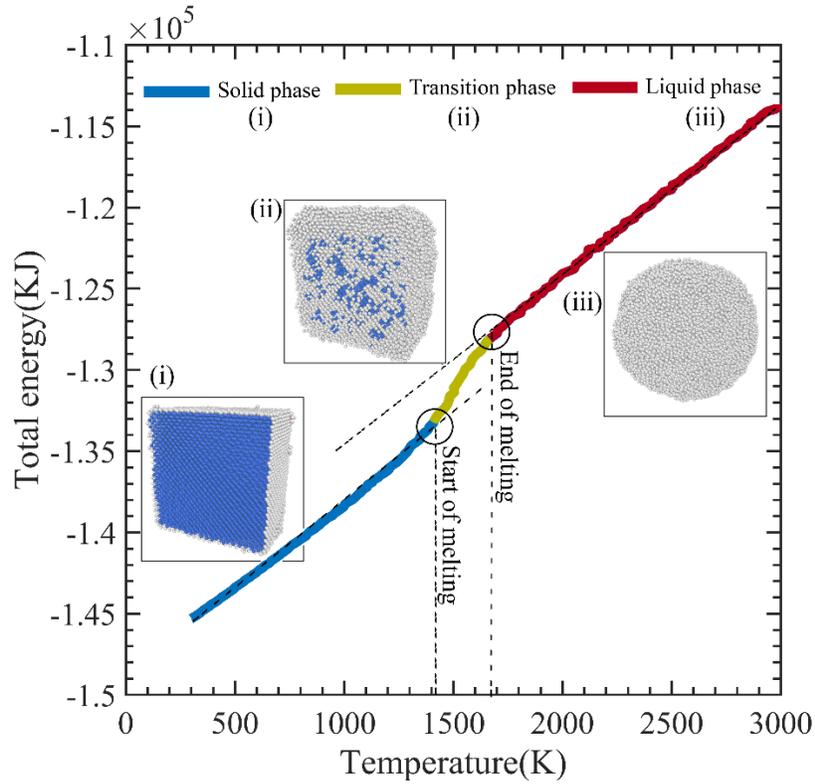

Figure 3: Change in total energy of cube as temperature increase 300K to 3000K and classification of the different phases of the material. Snapshot of cross-section of cube with 34461 atoms at (i)300K (ii)1478K (iii) 1705K to indicate its corresponding state at that temperature.

## 4. Results and Discussions

*4.1. Impact of size and temperature*

Figure 4(a) shows the stress-strain response of Inconel-718 NWs with different cross-sectional widths at 300K temperature, exhibiting the size effect. On the other hand, figure 4(c) shows the stress-strain relation of 3.17nm nanowire for different temperatures ranging from 100K to 800K. In both cases, the applied strain rate is $10^9$ $s^{-1}$.

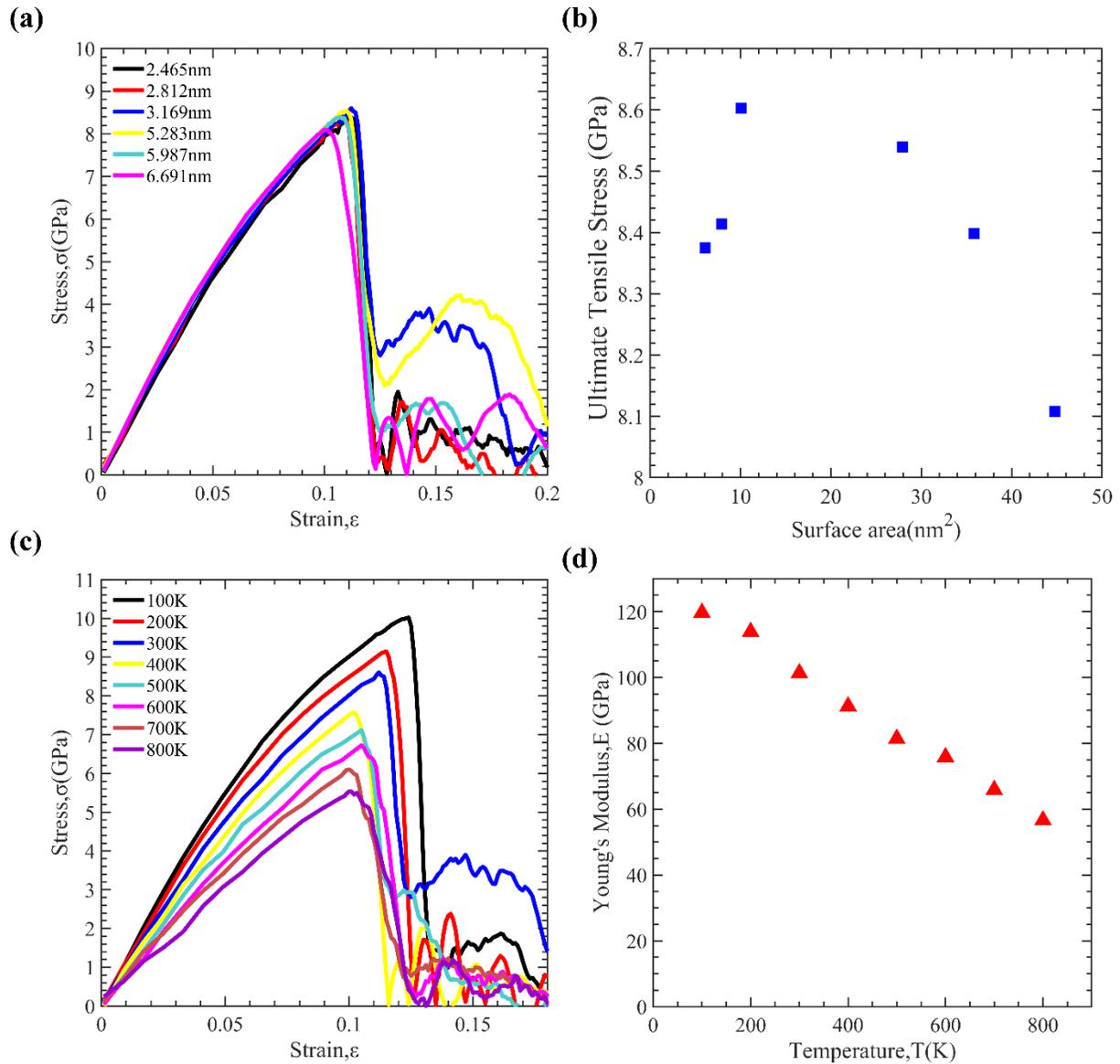

Figure 4: (a) Stress-strain curves for Inconel-718 nanowire as a function of nanowire width. (b)Variation of UTS of the nanowires with surface area(nm$^2$). (c) Stress-strain curves for nanowire of width 3.17nm as a function of temperature. (d) Variation of Young's modulus of the nanowire of width 3.17nm with temperature.

It is clear from the graph that all of the stress-strain curves for tension follow almost the same path until fracture. Due to its ductile nature[53], the nanowires do not fail catastrophically. A closer look at figure 4(a) reveals that the stress first increases and reaches to the ultimate tensile stress (UTS). Then it consecutively experiences a steep drop after which fluctuations of stress is observed due to flow stress with further increase in strain. Moreover, it is observed from figure 4(b) that with the increase of nanowire surface area UTS at first increases up to a maximum value of 8.60GPa and after which nanowires (with the larger surface area) exhibit a lower UTS. The strength of nanowires increases up to a surface area of 10nm² and any further increase in the surface area of the nanowire will result in lower strength. This behavior of the nanowires with the increasing size is in perfect agreement to the study done by Li et al[54].

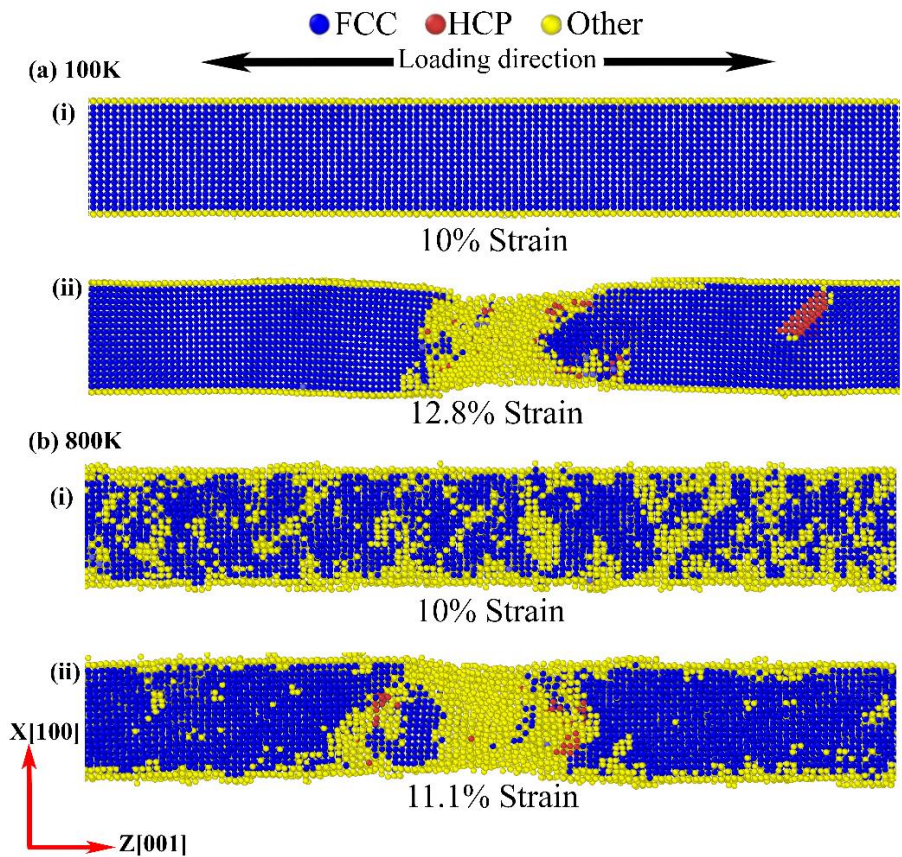

Figure 5: Sectional view of a portion of the nanowire of 3.17nm at (a) 100K temperature at (i)10% and (ii)12.8% strain and (b) 800K temperature at (i)10% and (ii)11.1% strain.

In contrast to the negligible effects of size on Young's modulus (E) and ultimate tensile strength (UTS), it is seen in figures 4(c) and 4(d) that the gradual increase of temperature drastically affects the E and UTS of the nanowire. The maximum value for E and UTS of the nanowire at 100K are 119.68GPa and 10.02GPa respectively. These values decrease almost linearly to 56.66GPa and 5.54GPa respectively at 800K temperature, which are about 50% of the values at 100K temperature. The physical phenomenon behind this is clearly illustrated in figure 5. From figure 5(b), it is clear that with the increase in temperature, the structure slowly changes from FCC to HCP and other unit cells. This occurs as a result of increased energy in the lattice due to higher temperature which makes the structure weaker. In figure 5(a), it is seen that at a lower temperature the atoms are less likely to be converted to hcp and other unit cells.

*4.2. Effects of strain rate and role of dislocations on material failure*

The influence of strain rate on the mechanical properties of Inconel-718 NW is depicted in figure 6 (a). Stress-strain curves of a NW with 3.17 nm width, at 300 K are shown for strain rates ranging from $10^8$ s$^{-1}$ to $10^{10}$ s$^{-1}$. It is observed from the figure that the fracture strength and strain decrease with the decrement of strain rate. It is due to the fact that at lower strain rate $10^8$ s$^{-1}$, the nanowire does not undergo any major dislocations, in comparison to other cases, as shown in figures 6 (b) and 7 (a). This phenomenon occurs because slip can nucleate easily at lower strain rates which causes the material to have a lower UTS. Hence, failure of material causes at a low strain. On the other hand, at a higher strain rate of $10^{10}$ s$^{-1}$, the structure undergoes significant dislocations as depicted in figure 6(b) and 7(c). Close observation reveals that initial dislocation induces strain hardening as a result material strength increases significantly resulting higher UTS under tension. The major type of dislocations is Shockley dislocation as observed in figure 7. When, dislocation density increases to a critical value (figure 6(b)), it causes a significant abrupt fall in the stress (figure 6(a)). At this point the slip occurs so the material starts to flow and dislocation becomes annihilated.

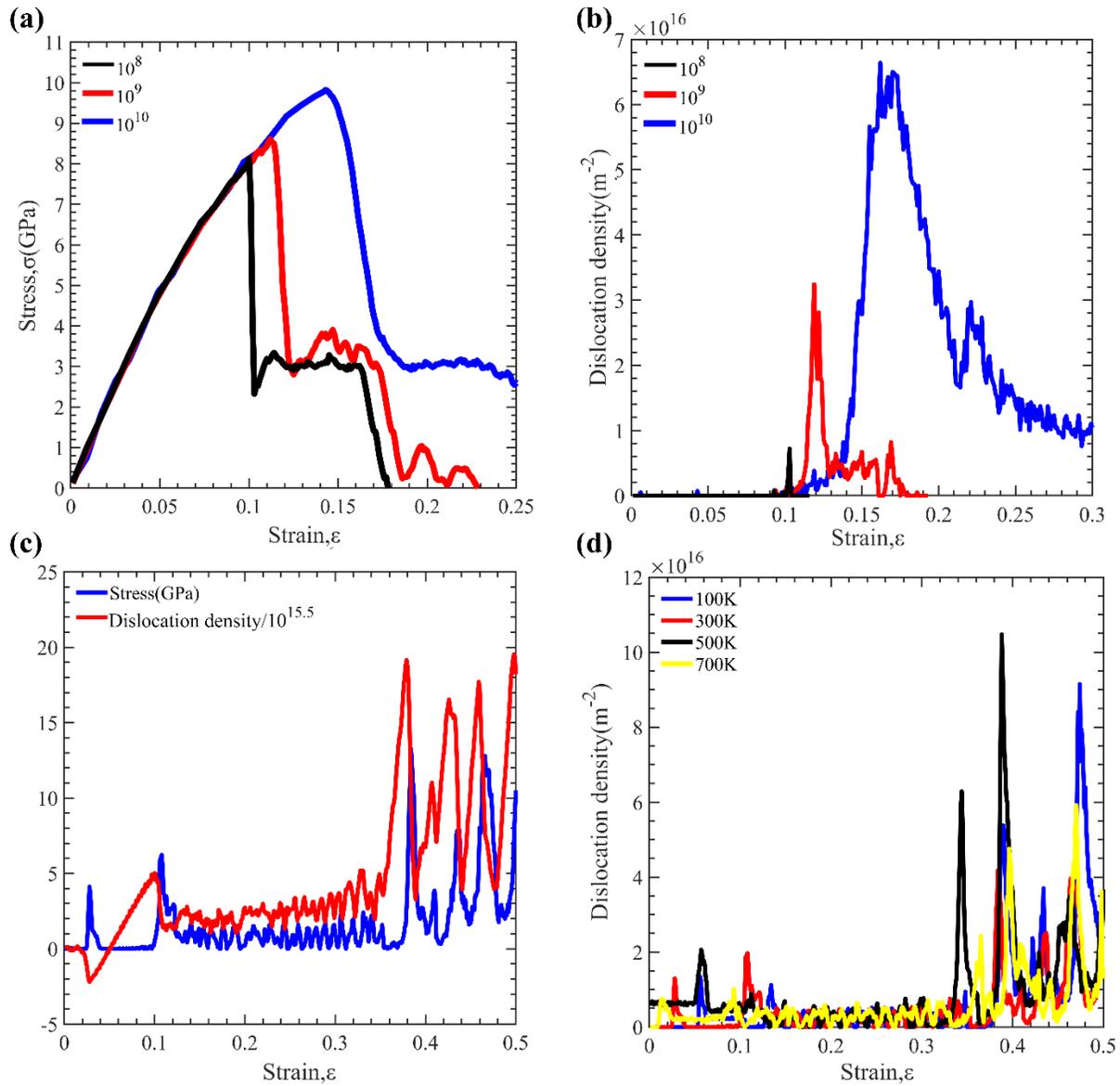

Figure 6: (a)Variation of Stress (GPa) for different strain rates, and (b) dislocation density with strain. (c) Visualization of relation between Stress (GPa) and scaled dislocation density with strain during compression of nanopillar. (d) Variation of dislocation density of nanopillar with strain at different temperature.

On the other hand, in case of compressive loading of nanopillar, this fluctuation is much higher and several peaks are seen as shown in figure 6(c). This is due to the activation of innumerable cross slip planes after yielding of material. Similar results are observed by Mojumder et. al.[55] for Al-Cu nano-pillar. From figure 6(c), it can be seen that there is an initial negative spike in the stress, this is due to fact that the nanopillar coming in contact with the boundary after relaxation and facing a compressive stress along the length of the nanopillar. An initial dip due to previous relaxation of all free surfaces of the nanopillar the stress slowly rose and reached the first yielding point. After that fluctuations start in the form of flow stresses. As the dislocations change their slip planes and interact, more fluctuations are observed.

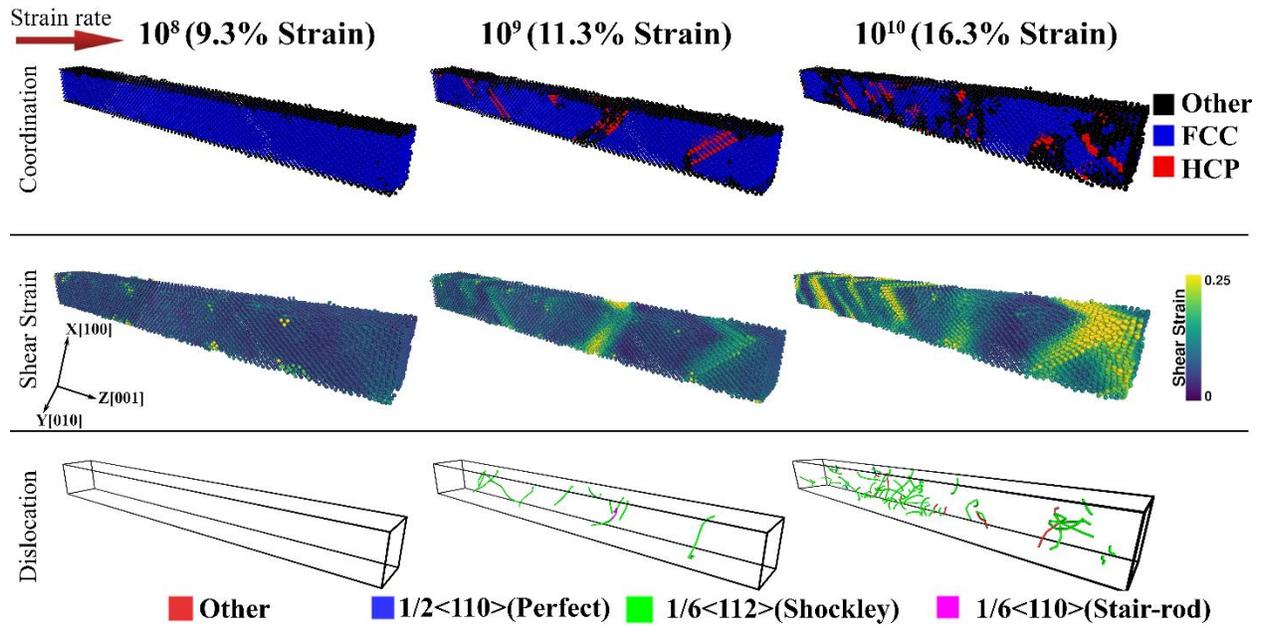

Figure 7: A visual representation of structural changes, shear strain and dislocations at different strain rates close to their fracture strain of 3.17nm nanowire at 300K. Green lines are Shockley type, blue lines are perfect, pink lines are stair rod type, and red lines are other types of dislocation obtained by DXA [56].

The stress-strain curve can further be explained by solid solution strengthening, as the lattice constants of Cr and Fe are 2.91 A° and 2.87 A° respectively which is smaller than that of Ni, 3.522 A°. As a result, when Ni atoms are replaced by Cu and Fe atoms, the lattice misfit creates a compression stress field inside the lattice. Therefore, when a load is applied on the lattice the dislocation tries to move along the slip plane and faces a strong energy barrier in its path. As a result, the dislocation chooses an energetically favorable path. A similar effect was observed by Lubarda et al. [57] on binary compounds. After an increase in stress the structure experienced major dislocation which caused the stress to subside. This phenomenon can be better understood by plotting the stress and favorably scaled dislocation density in the same graph (figure-6(c)). Evolution of dislocation density is same under loading for different temperatures as shown in figure 6(d). Initial dislocation density might change depending on the process history.

*4.3. Failure mechanism*

The Inconel-718 nanowire in the present study fails by shear in the {111} plane. In figure 8, the failure of a nanowire with a cross-sectional width of 3.17nm is shown at a temperature of 300 K with the aid of the shear strain parameter. Lower values of shear strain represent a perfect lattice, while higher values indicate a local defect of surface atoms. It is observed from figure 8(a) that the bond length of the nanowire increases uniformly up to a strain of around 10% and at this time the structure remains almost perfect with no major slip. At 12.8% strain, slip induced sliding is initiated at the nanowire surface which results in yielding and necking starts to appear in the NW. As observed in figure 8(c), local thinning occurs in a strain region of 12.8% to 22.6% and at a strain of 22.6%, the material experiences complete failure. To identify the slip plane, displacement vector analysis the nanowire is used which is shown in figure 8(b). It is observed that the slip propagates along a plane making 45° angle with the nanowire axis. The 45° angles of the shear planes with nanowire axes both from x-y and x-z planes represent shear along the {111} plane. Since the bond length along these planes is maximum the bond energy along this direction is minimum as a result slip induced shear occurs in {111} planes.

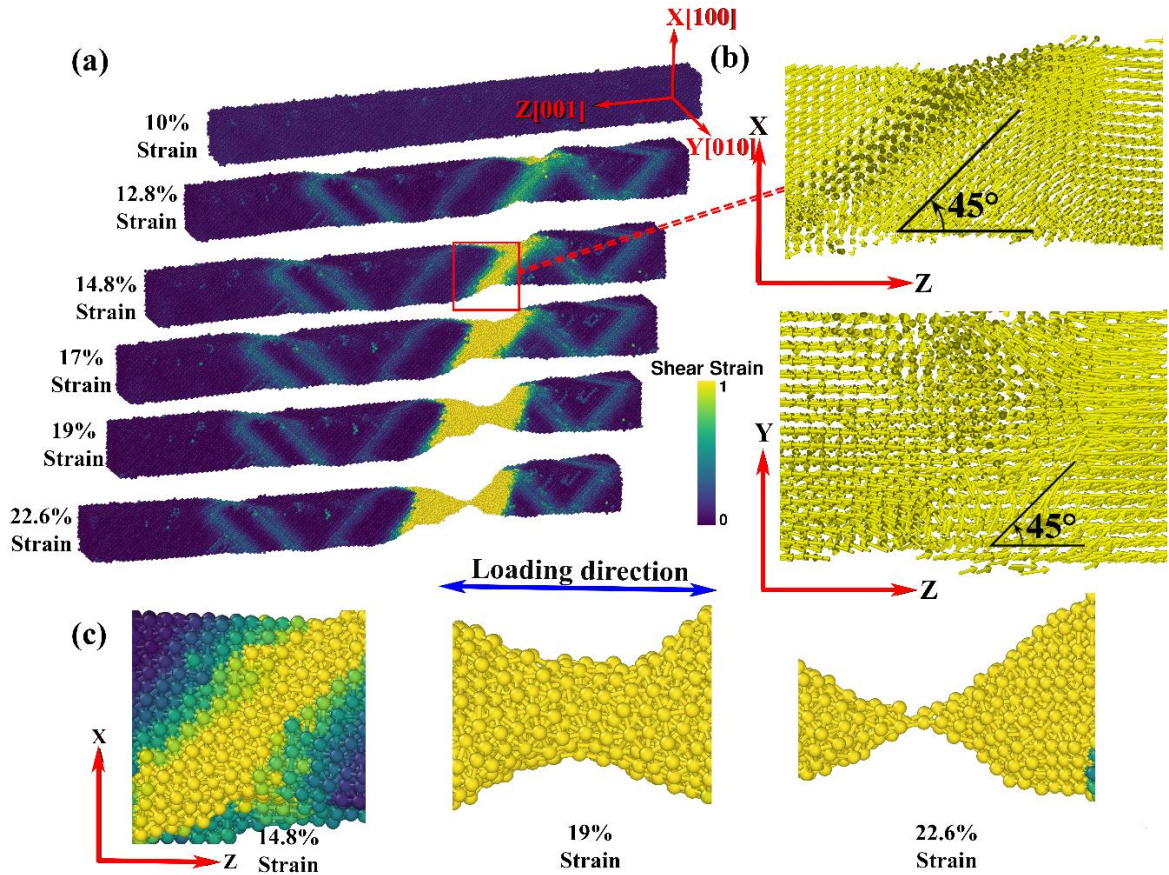

Figure 8:(a) Failure mechanism of Inconel-718 NW with a cross-sectional width (D) of 3.17nm at 300 K. (b) the atomic displacement values from different perspectives at 14.80% strain. (c) Bond breaking and necking under tensile loading for various strain level. Yellow color indicates bonds that are about to break under tensile loading. Colors of the atoms in (a) and (c) correspond to the indicated shear strain parameter values. Nanowire axes are provided in each column to guide the reader.

*4.4. Effects of cooling rates*

Figure 9(a) displays stress-strain curve for different cooling rates. The cooling rate plays a key role in affecting the microstructure phase during solidification. Based on the previous work, the cooling rate is divided into the conventional die casting ($10 - 10^3$ K/s), melt spinning ($10^5 - 10^6$ K/s)[58], liquid splat-quenching ($10^9 - 10^{10}$ K/s)[59], and pulsed laser quenching ($10^{12} - 10^{13}$ K/s) [60,61]. Here, the cooling rate in the range from $0.5 \times 10^{10}$ K/s to $1 \times 10^{14}$ K/s [60,61] is considered, due to the limit of the computational ability [61–64] and our interest to establish that molecular simulation can be a reliable method to observe the physics.

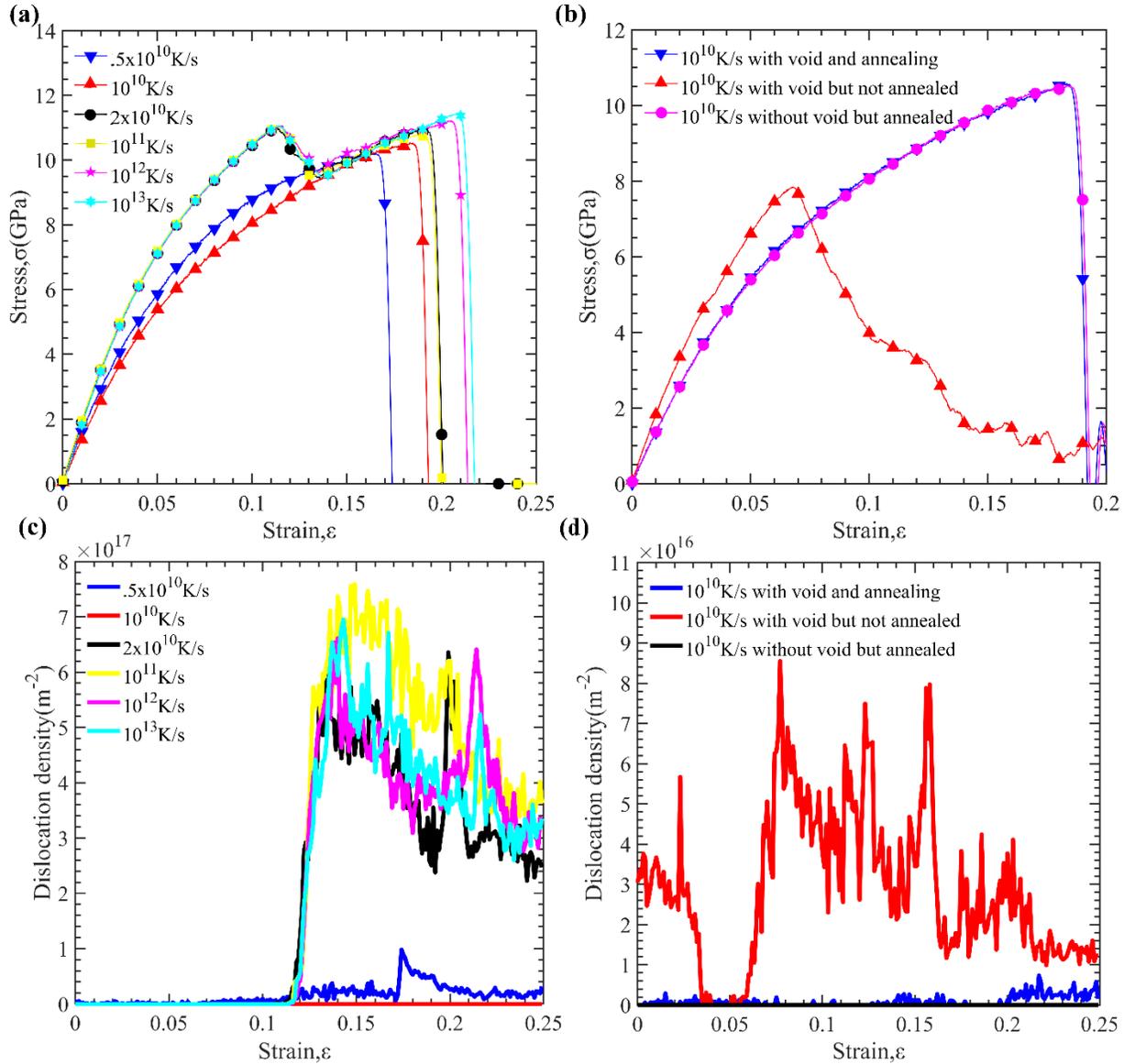

Figure 9: Variation of tress (GPa) with strain (a) for different cooling rates and (b) for different combinations of void and cooling rate. Variation of dislocation density with strain (c) for different cooling rates and (d) for different combinations of void and cooling rate.

It is clear that with the increment of cooling rate both material strength and fracture strain increase, and similar kind of result was observed for laser deposited Inconel-718 [65]. Interestingly from figure 9(a), it can be observed that at cooling rates lower than $10^{10}$ K/s, the material shows brittle type failure, while at cooling rates higher than $10^{10}$ K/s, ductile behavior is observed. This is due to that fact that at high cooling rates non-equilibrium structure (figure 10) is formed which leads to higher strength and fracture strain. However, at lower cooling rate, crystal structure is formed (figure 10) leading to lower strength and fracture strain like brittle material.

Figure 9(c) gives more insights to understand the fact with the help dislocation density. High dislocation density is observed for higher cooling rates which indicates ductile behavior. On the other hand, almost no dislocation density is found in case of lower cooling rates meaning brittle material. So, brittle to ductile transition occurs due to cooling rates.

Figure 9(b) presents stress-strain curves under different combinations of void and cooling rate. Here, we observe that without any cooling rate, when a central void is present, the material shows a significant lower strength. This phenomenon is better understood with figure 9(d) and figure 10, where it is clearly seen that without applying cooling rate, the material undergoes ductile failure due to dislocations. As stress concentration increases parallel to the direction of application of force, it results in greater amount of dislocations which eventually lead to earlier shear of the material. But, when a cooling rate of $10^{10}$ K/s is applied material exhibits improved strength and fracture strain irrespective of void. Cooling rates significantly affect the microstructure phase during solidification as shown in figure 10. when cooling rate of $10^{10}$ K/s is applied the whole FCC structure solidified to a new structure which is consists of other crystal structures and it shows improved properties and Brittle behavior. But, due to $10^{11}$ K/s cooling rate, a mixture of different crystal structures (figure 10) meaning a non-equilibrium structure is formed during solidification which results in ductile behavior. From figure 10, it is also visually clear that the non-equilibrium structure failure occurs due to dislocation which strengthen our explanation. Another surprising behavior is that geometry containing central void when subjected to cooling rates shows significantly higher strength which is almost same as annealed geometry without void (figure 9c). This is owing to the fact that during the solidification process the void is eliminated in the new structure which is healing mechanism as observed in figure 10.

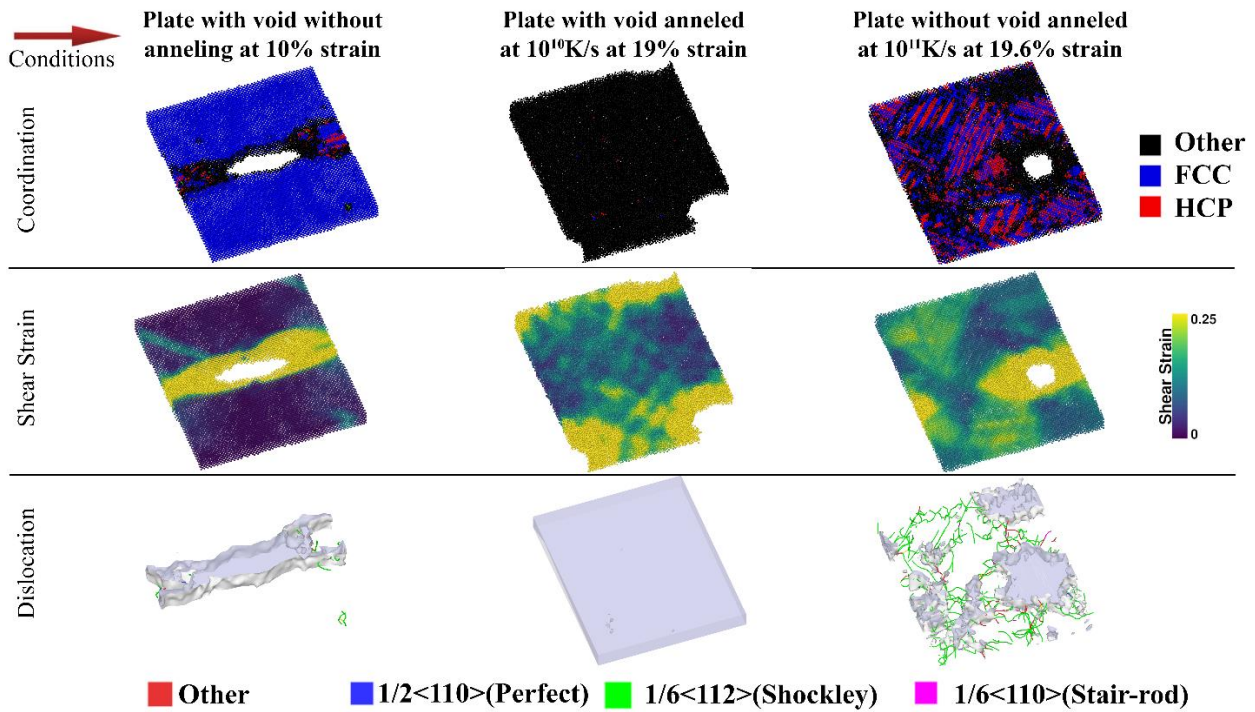

Figure 10: A Comparative study of the nanoplate at 300K with different combinations of cooling rate applied and presence of void. Green lines are Shockley type, blue lines are perfect, pink lines are stair rod type, and red lines are other types of dislocation obtained by DXA [56].

## 5. Conclusions

In this study, molecular dynamics is used to model, simulate, and provide insights into the deformation mechanism of nickel based super alloy Inconel-718 at the nanoscale. The study sheds light on how the physical conditions from manufacturing process can be incorporated into molecular dynamics study to observe nanoscale physics. Here, atomistic simulations have been carried out to investigate the role of dislocation density on material failure both under tension and compression. Practical application like how annealing can affect material strength is also observed in this study. Besides, effects of different temperature, size, and fracture mechanism of nanowire are also presented in this study. It has been observed that-

1) Strength of nanowire increases as the size of nanowire decreases initially when the surface area of nanowire along the direction of strain is greater than 10nm$^2$. After that any further decrease of the surface area causes the strength of the nanowire to decrease again.

2) With increasing temperature, the random vibration of the atoms in the nanowire increases and more atoms stray from their ideal lattice positions causing lower young's modulus and strength of the nanowire.

3) Increasing the strain rate allows the nanowires to experience more atomic dislocations initially causing strain hardening due to which the nanowires have a higher strength under tension. While Single peak in dislocation density is observed under tension, multiple peaks is observed under compression. During compression of a nanopillar the stress slowly rises and the pillar yields. After that the flow stress creates fluctuations which are created by increase in stress and then subsequent spike in dislocation which causes the stress to decrease again. Apart from this, temperature has no such impact on dislocation density.

4) A typical shear failure is observed for Inconel-718. In this case, dislocation causes extensive sliding on a {111} plane (tilted with respect to the NW axis), which leads to significant local thinning of the NW before fracture.

5) Cooling rates play a key role in material strengthening. It is observed that high cooling rates increase both material strength and fracture strain and low cooling rates do the opposite.

Surprisingly, cooling rates lower than $10^{10}$ K/s is applied to material, it shows brittle type failure and higher cooling rate greater than $10^{10}$ K/s helps material to exhibit ductile failure. In addition, void or porosity inside the material may change shape and impact on the depending on the cooling conditions.

## 5. Acknowledgements

The authors of this paper would like to acknowledge Multiscale Mechanical Modelling and Research Network (MMMRN) group for the technical support to conduct the research.